**Scales of influence on the settling velocities of synthetic, industrial and natural particles in grid turbulence**

Authors: Corrine Jacobs, Wilmot Merchant, Marek Jendrassak, Varavut Limpasuvan, Roi Gurka and Erin E. Hackett

*School of Coastal and Marine Systems Science, Coastal Carolina University, Conway, SC, 29528, US, (corresponding author) E-mail: ehackett@coastal.edu*

**Abstract**

The settling velocities of natural, synthetic, and industrial particles were measured in a grid turbulence facility using optical measurement techniques. Particle Image Velocimetry and 2D Particle Tracking were used to measure the instantaneous velocities of the flow and the particle's trajectories simultaneously. We find that for particles examined in this study ($Re_p$ = 0.4 - 123), settling velocity is either enhanced or unchanged relative to stagnant flow for the range of investigated turbulence conditions. The smallest particles scaled best with a Kolmogorov-based Stokes number indicating the dissipative scales influence their dynamics. In contrast, the mid-sized particles scaled better with a Stokes number based on the integral time scale. The largest particles were largely unaffected by the flow conditions. Using Proper Orthogonal Decomposition (POD), the flow pattern scales are compared to particle trajectory curvature to complement results obtained through dimensional analysis using Stokes numbers. The smallest particles are found to have trajectories with curvatures of similar scale as the small flow scales (higher POD modes) whilst mid-sized particle trajectories had curvatures that were similar to the larger flow patterns (lower POD modes). The curvature trajectories of the largest particles did not correspond to any particular flow pattern scale suggesting that their trajectories were more random. These results provide experimental evidence of the "fast tracking" theory of settling velocity enhancement in turbulence and demonstrate that particles align themselves with flow scales in proportion to their size.

**Introduction**

The behavior of particles in turbulent flows is important to many scientific and engineering fields, such as meteorology, oceanography, sedimentology, hydraulic and civil engineering. Whether these particles are atmospheric aerosols, contaminants in engineered systems, or sediment transport in the ocean, the flow within which these particles reside is often turbulent. In conjunction with these applications, two primary questions arise: (1) how do the particles disperse in the fluid, and (2) how do they settle out? The latter question has been studied less extensively, and depends largely on the settling velocity of the particles. For example, the settling velocity controls the residence times of aerosol contaminants in the atmosphere (Maxey 1987), the rates of particle collisions (Good et al. 2014) within the flow medium, the downward flux of sediment mass in the ocean (Scully and Friedrichs 2007), and the suspended sediment concentration in the ocean water column (Ha and Maa 2010).

Studies of settling velocities in stagnant fluids have resulted in both theoretical and empirical predictive models, including natural particles (Rubey 1933; Dietrich 1981). In contrast, results of settling velocities in turbulent flows are less conclusive. Murray (1970) examined particle-settling velocities in grid-generated turbulence and found that settling velocities can be significantly reduced in turbulence. Nielsen (1993) performed similar experiments and found that settling velocities can either be reduced or increased depending on the turbulence intensity, which was estimated based-on the grid velocity amplitude. More recently, Kawanisi and Shiozaki (2008) found results consistent with Nielsen (1993) for extreme cases (i.e., for very low or very high turbulence levels), but in intermediate turbulence intensities ($\sigma_p/V_0 \sim 0.2$-$5$, where $\sigma_p$ is the standard deviation of particle vertical velocity and $V_0$ is the particle's terminal settling velocity) the increase or decrease in settling velocity was Stokes (St) number dependent with smaller St numbers having an increased settling speed. Hence, to date, our understanding of the effects of turbulence on settling properties is incomplete.

Several theories exist on how turbulence might impact the settling speed. Three of the four theories suggest mechanisms that would decrease settling velocity. Specifically, these mechanisms are vortex trapping, non-linear drag, and loitering (Nielsen 1993). Vortex trapping refers to particles that can become confined inside forced vortices as demonstrated experimentally by Tooby et al. (1977). Nielsen (1984; 1992) further demonstrated that vortex trapping could also occur in irrotational vortices, such as those generated by surface waves. The second mechanism of settling velocity reduction is attributed to non-linear drag effects. As demonstrated by Wang and Maxey (1993) and Good et al. (2014) using direct numerical simulation (DNS) of particles settling in turbulent flows, settling velocities diminish only when non-linear drag terms are included in the numerical scheme. Finally, as discussed by Nielsen (1993), the loitering effect occurs when particles "oversample" the upward flow relative to downward flow, resulting in a decrease in particle settling speed. DNS results of Good et al. (2014) indicate that anisotropic flows, where horizontal fluctuations are small relative to vertical fluctuations, can increase the likelihood of loitering effects as the particle has little means to adjust its horizontal position. In the extreme (i.e., no turbulence "structure"), the loitering effect is akin to a random walk. One of the subtleties that differentiates vortex trapping and loitering is that the former needs a persistent eddy structure.

Maxey and Corrsin (1986) showed that inertia (or density) can cause heavy particles to spiral outward - a phenomenon knows as "fast tracking". Through theoretical and numerical simulations of a random cellular flow field, this study provided evidence that the trajectories of settling particles tend to align along S-shaped curves that follow the right-side edges of clockwise vortices and the left-side of counterclockwise vortices. This result was confirmed with the numerical simulations of Maxey (1987), and is currently, to our knowledge, the only theory that explains the increased settling velocities observed in experimental data (e.g., Nielsen 1993; Yang and Shy 2003; Yang and Shy 2005; Kawanisi and Shiozaki 2008).

Although the conceptual idea of "fast-tracking" has been around for some time as well as experimental evidence of increased settling velocities, few studies have attempted to connect turbulence patterns with the observed increased velocities. Yang and Shy (2003) conducted a set of carefully constructed experiments in an effort to validate numerical data (Wang and Maxey 1993; Yang and Lei 1998). Their results were consistent with the DNS results where they observed a maximum in the increased settling speeds at Kolomogorov-based Stokes numbers near 1: $St_K = \tau_p/\tau_K \approx 1$, where $\tau_K$ is the Kolomogorov time scale and $\tau_p$ is the particle relaxation time defined as $\rho_p d^2/18\mu$. Here, $\rho_p$ is the particle density, d is the particle diameter, and $\mu$ is dynamic viscosity. Furthermore, using wavelet transforms, they showed that at $St_K \approx 1$ the frequency of particle motion and fluid motion are similar and assert that under these conditions the relative slip velocity between the particles and the flow is smallest resulting in decreased drag and therefore larger settling speeds. In contrast, at other $St_K$ numbers, they did not find a correspondence between particle motion and fluid motion. We note however that the wavelet transform technique employed by that study could only discern integral time-scales and Taylor microscale time-scales of the fluid motion.

Through the use of DNS and large-eddy simulations (LES), Yang and Lei (1998) examined the connection between turbulent scales and enhanced settling velocities. They found that the relatively small-scale flow features, corresponding to the dissipation spectral peaks, largely governed the locations of the fast tracks. However, they also found that the large energetic eddies are important for increasing the settling velocities, while scales smaller than that associated with the dissipation spectral peak had no effect on the fast track locations or the increased settling speed. The accumulation of particles in the periphery of the small-scale vortices, where vorticity is comparatively low and drag is reduced. Although this drag reduction is necessary for an overall increase in settling speeds, the magnitude of the drag reduction for each individual particle is tied to the energy-containing scales since the difference between the local fluid velocity and the particle's velocity determines the drag. Yang and Lei (1998) further assert that the Kolomogorov time scale tends to collapse settling velocity data because is related (by a constant) to the scales associated with the peaks of the dissipation spectrum, rather than because the Kolomogorov scales impact the settling velocity directly. It appears that more experimental data is needed to decipher the connection between settling speed and turbulent features.

In an effort to investigate the role of turbulence in modifying particles' settling velocities, this study performs a set of particle-settling experiments in a range of near-isotropic turbulence conditions, generated by using an oscillating grid facility. Particle trajectories are computed through high-speed imaging of the descending particles. Simultaneous with these data, particle image velocimetry (PIV) data are acquired to enable the computation of flow's 2D spatial distributions. First, we examine how the turbulence impacts the settling speeds of various particle types (i.e., different densities, sizes, and shapes) that cover a range of particle Reynolds numbers between 0.4 - 123 and relative turbulence intensities of $V/v_\eta$ = 1.5 - 60 and $V/v'_{RMS}$ = 0.4 - 15, where V is the particle settling speed in turbulence, $v_\eta$ is the Kolomogorov velocity scale, and $v'_{RMS}$ is the root-mean-square of the turbulent fluctuations. Subsequently, we examine the turbulent scales of the fluid using proper orthogonal decomposition (Lumley 1970) and compare it to the shape of the particle's trajectories to relate the size of the turbulent structure to its impact on the particle's trajectory.

**Experiments and Methods**

Experiments were conducted in a grid turbulence facility, where multiple types of particles, ranging from natural to synthetic with different sizes and densities, have been investigated under various turbulence conditions.

*Grid Turbulence Facility*

As depicted in Fig. 1, a glass tank was fabricated with dimensions of 0.5×0.5 $m^2$ in cross section and 1 m in height. This tank contains a vertically oscillating grid of square bars with mesh size of 31.5 mm. Driven by a 0.75 kW variable speed electrical motor and an eccentric flywheel, the grid motion ranges 82 mm in peak-to-peak amplitude (stroke) and at a frequency of up to 10 Hz. In the present experiment, the grid was set to oscillate between 2 - 7 Hz and the center of the vertical oscillation was located at the center of the tank, i.e., y = 49.2 cm measured from the bottom of the tank. Therefore, the corresponding grid turbulent Reynolds number, $Re_g$ = fSM/ν (where f is the grid frequency, S is the stroke, M is the grid mesh size and ν is kinematic viscosity), varied between 5,166 and 18,000. The grid oscillated for 20 minutes before the start of each experiment to ensure that the flow had reached steady state conditions (thereby, approaching homogeneity) prior to the start of each experiment. Once mixed conditions were achieved, the particles were introduced uniformly at the top of the tank and descended 70 cm in the tank before reaching the measurement region. This procedure ensures that the initial conditions of the particle motion at the top of the tank dissipated and that the particles reached their terminal velocities. The PIV and high-speed imaging were subsequently initiated to collect data.

*Particle Characteristics*

Five types of particles were chosen for the characterization of the settling velocity under turbulent conditions: 3 industrial types, one natural and one synthetic. Table 1 presents a list of the particles and their characteristics. Having three different types of particles (natural, industrial and synthetic) permits us to address the interaction between the flow and the particle characteristics such as size, shape, and density. The natural sand (density of 1650 kg/m$^3$) was collected from the local beach at 38th Avenue North in Myrtle Beach (South Carolina) about 5 meters inland from the low tide line and was allowed to completely dry before experimentation. Sands used in industrial manufacturing processes were also examined; these sands have densities of 2640, 3970, and 2200 kg/m$^3$ (Conbraco Inc., personal communication). Additionally, synthetic hollow glass spheres (Potters Industries) with a density of 1440 kg/m$^3$ were employed. The synthetic spheres are almost perfectly spherical while the others are irregular in shape. The particles were initially classified by size using sieves; however, shape irregularities render the sieve mesh size to be an unreliable indicator of size. Thus, the mean particle diameter, $d_{50}$, for each sieve-based size class was measured using a laser diffraction particle analyzer. Fig. 2 shows a sample output from the laser diffraction particle analyzer along with sample images obtained under a microscope to highlight the differences in particle shape.

*Flow Field Measurements*

A Particle Image Velocimetry (PIV) system was used to measure the flow field in the tank. This system consists of an Nd:YAG dual head laser with 120 mJ/pulse operating at 15 Hz with a wavelength of 532 nm, optical lenses to form a vertical light sheet approximately 5 mm thick, a 2M (1600×1200 pixel$^2$) double exposure CCD camera with a dynamic range of 12 bits and a 50 mm lens. The timing between the laser pulses and the camera exposure were controlled using a synchronizer to obtain pairs of consecutive images. The flow was seeded with hollow glass sphere particles with a mean diameter of 11 μm, comparable to the turbulent scales of the flow (Melling 1997). As shown in Fig. 1, the camera was placed 54.5 cm from the tank to capture a field of view (FOV) of 14.2×10.3 cm$^2$.

A total of 500 pairs of images per experiment were acquired in order to obtain convergence of statistical properties. PIV images were correlated using cross-correlation analysis with an interrogation window of 64×64 pixels and 50% overlap. The PIV data were subsequently filtered to remove outliers using a 3 standard deviation local and global filter, resulting in ~5% erroneous flow vectors that were replaced with interpolated vectors, per velocity map.

*Particle Tracking Image Processing*

To estimate the particle velocity, we use a 2D particle tracking method. The tracking is based on acquiring time resolved images with a high-speed CMOS camera using a 50-mm lens located 62 cm from the tank, forming a FOV of 23.2×23.2 cm$^2$ that overlapped with the PIV FOV as shown in Fig. 1. The CMOS camera has a spatial resolution of 1000×1000 pixel$^2$ operating at up to 1 kHz. In our experiments, the camera operated at 60 frames per second due to the particle's small relative velocities. The tank was illuminated from underneath by a continuous monochromatic (532 nm) LED line light (~2 cm thick). The backscattering of the LED light by the particles was recorded using the CMOS camera. Each run recorded about 2,500 images of settling particles in the tank, resulting with an average tracking of several seconds per descending particle. Particle concentration in the tank was chosen to ensure that individual particles could be tracked. On average, about 25 particles appeared in each image, which is sparse enough to track the particles. No particle flocculation was observed. Thus, tracking results are assumed to be valid only for settling phenomena not influenced by high particle concentration.

The particle-tracking algorithm contained the following steps. First, images were converted to black and white using a gray scale threshold, which varied from 300-1,500 depending on particles' reflectivity. This threshold was determined by observing the histograms of pixel intensities of the images within a run and examining the separation between the background gray level and the peak associated with particle reflections. Second, a blob analysis was performed that required a connectivity of 8 pixels in order to identify an object. The centroids of these objects (individual particles) were then determined, resulting in identification of the particle's position in each image (frame). Next, a search window was applied to the next consecutive image. The search window around identified particles in the first image looked for the same particle in the second image, and so on. The size of the search window was chosen to be ±20 pixels in the x direction, -30 pixels and +20 pixels in the y direction based on an optimization process. The window size was at least 3 times the largest/fastest particle's displacement in stagnant flow. The difference in length of the search window between x and y is due to gravity; we expected that the displacement of the particles between images will be larger in the y direction than the x direction. If multiple particles were found within the search window then the particle track was terminated. This procedure increased the confidence that the particle identified in the prior image was the same one in the subsequent image. The particles were tracked until they were no longer found inside the search window of the next image. Once all particles were found in consecutive images, the data was stored together as particle trajectories over time, as seen in Fig. 3 for one experiment. Because the measurements of the flow field and the particles were done simultaneously (with the laser and LED illumination sources, respectively), overexposed CMOS images due to laser pulsing from the PIV were removed from the original time series of the CMOS image data based on a cumulative intensity threshold. The particle's 3D trajectory is imaged onto a 2D plane; however, errors associated with significant out-of-the-plane motion are minimized by the experimental setup and analysis. Because the LED light sheet is only 2 cm thick and long particle trajectories are used in the analysis (see Results section), the tracked falling particles tend to be on the 2D plane illuminated within the 2 cm light sheet.

**Results**

The impact of turbulence on particle settling velocities is examined using the non-dimensional Stokes parameter. Here, we combine the results obtained by the PIV technique and the 2D particle-tracking algorithm, which enables us to couple the flow field data and the particle trajectory analysis. Throughout the analysis, the different particles are clustered into 13 groups depending on particle characteristics of type, size, and density.

*Flow Field Characterization*

PIV is used to obtain instantaneous two-dimensional velocity measurements. As illustrated in Fig. 4, the mixing created by the oscillating grid yields conditions close to homogeneous flow with zero mean flow in any direction, which is further supported by Fig. 5, which shows the root mean square (RMS) of the velocity fluctuations in the x direction, $u'_{RMS}$, versus height with respect to the bottom of the tank. Increasing the frequency of the oscillating grid results in an increase of the RMS velocity as depicted in Fig. 5. The RMS values increase from about 0.4 up to 2 cm/s over a range of frequencies between 2 to 7 Hz, illustrating a linear trend between the external forcing and the turbulence (shown in Fig. 6), which is in agreement with the results of Shy et al. (1997). The RMS velocity profiles in Fig. 5 are nearly homogenous in the vertical direction, which suggests well-mixed conditions in the experimental set-up. These well-mixed conditions in our experimental set-up allow for the comparison of settling velocities with varying turbulent conditions.

To characterize the coupling between the flow field and the settling particles, the flow scales have to be estimated and used for scaling the phenomena, such as the integral scale, $\mathcal{L}$. The integral scale is defined here by the auto-correlation function of the vertical velocity component (v) obtained from the PIV in the transverse direction (x) (see Fig. 7). Specifically, the area under the curve (integral) between the origin and first zero-crossing of the auto-correlation function is used to estimate the transverse length scales (Pope 2000). Table 2 presents the values of the integral scales as a function of oscillating grid frequency. The variation of the integral scale with frequency shows that, although the turbulence intensity (as reflected by the RMS values) increases with the increase of external forcing (Fig. 6), the scale is largely independent of the frequency for the range of 2 - 7 Hz. This behavior is consistent with Shy et al. (1997) who found that the integral time scale is inversely proportional to the grid-oscillation velocity whilst the turbulence intensity is proportional to it.

Next we estimate the dissipative scales. These estimates are used to compare non-dimensional numbers based on these scales to those based-on the integral scales. The average rate of dissipation is estimated using the relationship $\epsilon \approx v'^3_{RMS}/\mathcal{L}$ (Kolmogorov, 1941). The Kolmogorov length is defined as $\eta = (\nu^3/\epsilon)^{1/4}$, with $\nu$ as the kinematic viscosity of the fluid, the Kolmogorov time scale is $\tau_k = (\nu/\epsilon)^{1/2}$, and the Kolmogorov velocity is $u_\eta = (\epsilon\nu)^{1/4}$. Table 2 depicts the Kolmogorov scales for the varying grid frequencies. We note that $\eta$ does not vary much over the change in frequencies and is of the same order of magnitude (~ tens of microns). These values are comparable with Yang and Shy (2003) where similar flow conditions were obtained. Both the integral length scales and the Kolmogorov length scales are used in our evaluation of the effect of the turbulence on the particle settling velocities.

*Particle Trajectories*

Particle tracking is used to determine the kinematic behavior, mainly settling velocities, of the particles (see Table 1) when introduced to turbulence. Following the experimental procedure as described in the section on *particle tracking image processing*, we have analyzed the high-speed images to obtain the 2D particle trajectories as they settle in the tank. These trajectories are decomposed into horizontal and vertical components over time. These decomposed trajectories are first smoothed using a 9 point (0.15 seconds) zero phase-shift boxcar filter (i.e., two-way filtering). The particles' instantaneous velocity at each point along their trajectories was subsequently calculated using ($2^{nd}$ order) forward differentiation. Samples of the instantaneous velocity distributions (horizontal and vertical) are shown in Fig. 8. We define the mean of these vertical velocity distributions to be the particle's "settling velocity". The longest observed trajectories for each run (particle group) were used in this analysis and the histograms represent no less than 2,500 particle velocity samples. This analysis is performed in both stagnant and turbulent flow.

Fig. 8 demonstrates how the distribution of the particle's velocity changes as the frequency increases from 0 - 7 Hz for the 261 μm natural sand and the 71 μm synthetic particles. The histograms of the particles' velocities exhibit a near Gaussian distribution, and the overlaid red curve represents the corresponding Gaussian probability distribution function (PDF). Clearly, the variability of the settling velocity (width of the PDF) increases with increasing frequency, demonstrating the relationship between increasing variability of settling velocity and the increasing turbulent intensities. This variability is characterized in the results that follow by the standard deviation of these distributions. The mean transverse velocity component, u, was approximately zero as expected because there is no mean flow in the tank.

We also measured the stagnant flow settling velocity, which is used as a reference and as a scaling factor for the settling velocity results obtained in turbulent conditions. As one of the seminal studies on this topic, Dietrich (1982) developed an empirical equation that accounts for the effects of size, density, shape, and roundness on the settling velocity of natural sediment. We used these equations to validate our methodology for computing the settling velocities. The filled circle markers in Fig. 9 represent the results for stagnant flow, which align well with Dietrich's empirical curves. We do not quantitatively consider the shape of particles in this comparison (i.e., the Corey Shape Factor or CSF), but we know the particles have various shapes (see Fig. 2). Thus, we show the Dietrich (1982) curves for various CSF in Fig. 9.

The settling velocities in turbulent conditions are compared to those in stagnant flow. As seen in Fig. 9, a majority of the particles show enhancement in settling velocities in turbulent conditions; however, the intensity of this enhancement varies greatly. While the figure suggests a trend in particle behavior, it is insufficient to demonstrate how the turbulence modifies the particle's settling velocity. The quantities and scales discussed in the flow field characteristics sections are used to characterize how the settling velocity changes with turbulent conditions as well as to examine how turbulent scales impact the settling velocity. These results are discussed in the next subsection.

*Influence of Turbulence on Settling Velocities*

Here, we combine the results obtained by the PIV technique and those from the 2D particle tracking. This combination enables us to examine the effects of turbulence on the settling velocity and how spatial flow patterns relate to particle movement.

*Modification of Settling Velocity due to Turbulence*

To examine the effects of turbulence on settling velocity, the corresponding behavior of particles suspended in a fluid flow is typically characterized by the Stokes number (Kawanisi and Shiozaki 2008; Good et al. 2014). The Stokes number is the ratio between the particle response time and a characteristic timescale of the flow. If the characteristic fluid time scale is based on the integral length scale, then the fluid timescale is defined as $\tau_f$, which is $\mathcal{L}/v'_{RMS}$. Using this definition for the fluid time scale, the Stokes numbers ranged from $St_l = 7 \times 10^{-5}$ to $10^{-1}$ as seen in Fig. 10 for the various particles and turbulence conditions tested. Fig. 10 depicts the Stokes number versus the settling velocity normalized by their settling velocity in stagnant flow. The normalization highlights whether the particles experience enhancement or reduction during settling, where $V/V_0 > 1$ indicates enhancement and $V/V_0 < 1$ a reduction. As shown in Fig. 10 there are few instances that showed reduction in settling velocity with only one showing a rather significant reduction at the lowest Stokes number. The majority of the particles have enhanced settling velocities. Yet, some of the particles, particularly the large ones ($St_l > 10^{-2}$), exhibit nearly no change.

We attempted to divide the particles behavior into three groups based on a range of Stokes number. The particles with large Stokes number ranging from $10^{-2}$ to $10^{-1}$ show essentially no enhancement. These correspond to particles, both natural and industrial, that have a diameter larger than 500 μm. Because these particles have $St_l$ numbers approaching 1, the particle relaxation time and the fluid time scale are similar; thus, the particle cannot respond fast enough to the fluid motion. For Stokes number range of $10^{-3}$ to $10^{-2}$, the particles primarily show a linear trend in Fig. 10 of the settling velocity with increasing Stokes number. For a fixed particle type (i.e., constant $\tau_p$), this trend means that the particles experience an exponential enhancement of their settling velocity as the integral time-scale decreases (or equivalently, as the v'$_{RMS}$ increases). The particle sizes that fall in this group ranges from 150 - 500 μm. Particles with Stokes numbers less than $10^{-3}$ demonstrate high variability, and there is no clear trend with respect to Stokes number, although large enhancement is observed.

Fig. 11 shows the variability of the settling velocity as a function of $St_l$. The standard deviation of the settling velocity is normalized by the standard deviation in stagnant flow. Clearly, the variability of the settling velocity increases as $St_l$ decreases. However, for a fixed particle relaxation time ($\tau_p$), the variability of the settling velocity generally linearly increases (in log space) as the fluid time scale decreases (v'$_{RMS}$ increases) for all groups, although the slope is not consistent.

The variation (or scatter) of the particle settling velocities for the small particles shown in Fig. 10 suggests that a different fluid time scale may be more relevant for these particles. Prior research has demonstrated that maximum enhancement occurs at $St_k \sim 1$, where $St_k = \tau_p/\tau_f$, where $\tau_f = \eta/v_\eta$. Fig. 12 depicts the settling velocity of the particles normalized by the settling velocity in stagnant flow as a function of the Stokes number based on Kolomogorov scales. We observe maximum enhancement in the range of $St_k = 10^{-2}$ to $10^{-}$ [1], which is slightly lower than that found by Yang and Shy (2003). Whilst the large particles exhibit no change in settling velocity as in Fig. 10, the small particles present a more linear relationship between $St_k$ and the normalized settling velocity in comparison to Fig. 10. This trend appears to be valid also for the mid-sized particles with a smaller slope although not as pronounced as it was in Fig. 10.

We conclude that large $St_l$ particles are mainly driven by their own inertia and are not able to respond to the flow; consequently turbulence plays a secondary role in their motion. The mid-size $St_l$ particles seem to be influenced more by the integral scales of the flow (in comparison to the smallest particles), which are the turbulent scales that are coupled with the external forcing applied to the grid. Finally, the small $St_l$ particles are most influenced by the dissipation scales of the flow. Even though the flow is not in the high Reynolds number range, and presumably a separation of scales does not occur to allow a clear differentiation between large, intermediate and small scales, the particles appear to be influenced by different scales of the fluid motion.

*Role of Turbulent Scales in Enhancement of Settling Velocities*

Prior studies suggest particles interact with the underlying turbulence and tend to favor certain regions of the flow. Along their trajectories, particles interact with vortices and the crossing trajectories cause the particle to be swept to the downward side of eddies (Aliseda et al. 2002; Nielsen 1993). Thus, while the turbulence also influences the drag experienced by the particle, the primary effect of the turbulence on a particle is to contribute to a net force leading to particle acceleration or deceleration. This reasoning suggests that a correlation should exist between the particles' trajectories and the turbulent scales of the flow, consistent with the results in the proceeding section.

To characterize the interaction between the particles' motion and the flow patterns, we apply proper orthogonal decomposition (POD) to the flow field (Lumley 1970) to estimate the spatial scales of flow patterns, and also estimate the radius of curvature for the particles' trajectories. Using both techniques will allow coupling of the particles' kinematics with the flow dynamics. The choice of using POD will enable a spatial description of the flow patterns in the grid facility in a statistical manner. The analysis complements and attempts to strengthen the conclusions drawn from the non-dimensional analysis presented above. POD is applied to the velocity fields (obtained from PIV).

Following a similar procedure as described in Gurka et al. (2006) and Taylor et al. (2013), the decomposition is performed using the snapshot method (Sirovich 1987). The turbulent scales are evaluated for each of the 13 particle groups at each of the 6 frequencies. The decomposition is performed on the entire data set and the velocity reconstruction (as well as flow pattern representation) is based on the first 50 modes. The presentation of the modes is done through calculating the vorticity from the reconstructed velocity field. We choose the first 12 modes as a means to characterize the flow features. Together, these modes represent more than 95% of the energy as reflected in the decomposition in which the entire data set is assumed to contain 100% of the energy. Every two consecutive modes were linearly combined due to mode-pair symmetries, resulting in six mode pairs to examine the vortical structures. To obtain a quantitative measure of the flow features, the width and height of all vortical patterns were estimated and averaged to obtain the scale of the vortices in each mode pair (see Fig. 13). The "edge" of the vortex was defined using a threshold of 95% vorticity relative to the maximum.

The particle's kinematics were characterized through the calculation of the curvature radius of the particle trajectories using:

$$R = \frac{\left[1+\left(\frac{dy}{dx}\right)^2\right]^{3/2}}{\left|\frac{d^2y}{dx^2}\right|} \qquad (1)$$

where the particle's horizontal and vertical positions are x and y respectively. A sample of the results is presented in Fig. 14, which presents the histogram of the curvature radius of all particle trajectories in one particle group (i.e., type and size) for a given grid frequency. The distribution means are used for the nominal radius of curvature for each particle group and flow condition.

The radius of curvature is matched with the vortices' scales for the different modes, as shown in Fig. 15. This figure depicts the relationship between the mean curvature radius of the particles' trajectories and an averaged size of the vortical patterns as a function of mode numbers. The figure also demonstrates how the trajectories of the particles align themselves with the vortices of different scales within the flow. The mid-range particles that exhibited a more coherent trend with $St_l$, appear to be influenced by vortical patterns associated with modes 3 - 6. Note that the range of integral length scales is highlighted by the vertical dotted lines, which is near the range most of the mode 3 - 6 vortices sizes fall. Following the conceptual model outlined by Nielsen (1993), this result suggests that the observed enhancement in the settling velocity of these particles is associated with the particles traveling on the downward side of eddies of this size range. The large particles which seem to correspond best to the lowest mode numbers do not seem to have trajectories that connect well to the flow patterns. This result suggests that they do not adhere to the Nielsen (1993) fast track enhancement model, which may explain why they do not exhibit any enhancement. Their inability to be responsive to the flow indicated by the large scatter between the vortex size and curvature radius of the particles' trajectories is consistent with their $St_l$ values being near one; consequently, they are more subject to the loitering effect as described by Nielsen (1993). This loitering effect is likely responsible for the slight enhancement observed by these particles rather than fast tracking. The small sized particles appear to follow the lower energy eddies, as depicted in the higher modes (> 6), which corresponds to the smallest scales of the flow. This result is consistent with the finding in the previous section that these particles' settling velocity scales more coherently with $St_k$. These results also support the numerical data of Yang and Lei (1998). In that study (based on slightly larger particles), the authors predicted that different scales influence different sized particles while our results indicate that this trend extends to smaller scales. Regardless, our results are generally consistent with their finding that both the dissipative scales and energy containing scales can influence the enhancement of the particles' settling velocity depending on particle size.

**Conclusions**

The settling velocities of particles were examined under varying turbulence conditions. The parametric study was performed using particles of different sizes ranging from 70 – 1400 μm and densities ranging from 1400 - 3900 kg/m$^3$ in an oscillating grid facility. The turbulence levels of the flow conditions were varied by changing the grid frequency resulting in different levels of mixing in the tank. Two methods (PIV and 2D particle tracking) were used to measure the flow field velocities and the particles' trajectories simultaneously.

The obtained results were presented in terms of Stokes number, using both the integral time scale and Kolomogorov time scale, and the normalized settling velocity (with respect to a stagnant flow condition). For the particles and turbulence intensities examined in this study, we find that particle settling velocity is primarily either unchanged or enhanced relative to stagnant flow. The smallest particles scaled best with the Kolmogorov-based Stokes number indicating that they are influenced more by the dissipative scales, consistent with the numerical results of Yang and Lei (1998). In contrast, the mid-sized particles scaled better with the Stokes number based on the integral time scale. The largest particles did not follow any scaling and were largely unaffected by the flow conditions. Maximum enhancement in the settling velocity was observed for $10^{-4} < St_l < 10^{-3}$ and at $10^{-1} < St_k < 10^{-2}$, where the latter is similar to the experimental findings of Yang and Shy (2003).

These Stokes number results were further supported by a POD statistical analysis, which was used to associate the particles' trajectories with turbulent scales. More specifically, we examined the spatial flow patterns that govern the particles' dynamics by applying POD to the velocity field and creating a reduced-order vorticity reconstruction, and comparing it with the estimated radius of curvature of the particles' trajectories. The small particles were found to have trajectories with curvatures of similar scale as the small flow scales (higher POD modes) while mid-sized particle trajectories had curvatures that were similar in size to the larger flow patterns (lower POD modes). The curvature trajectories of the largest particles did not correspond well to any particular flow pattern scale suggesting that their trajectories were more random, similar to the loitering effect described by Neilsen (1993). The correspondence of the mid-sized and smallest particles to different flow patterns as identified with the POD analysis suggests that they do preferentially align themselves with flow patterns consistent with the fast tracking conceptual model discussed in Neilsen (1993). The particles tend to align themselves with flow scales in proportion to their size.

**Acknowledgements**

The authors would like to thank the II-VI Foundation for their support of this effort. The authors would also like to thank Dr. Jenna Hill for use of the laser diffraction size analyzer and Mr. James Sturgeon at Conbraco Inc. for donating the industrial particles. The support of Coastal Carolina University's Provost office through the Academic Enhancement Grant and the Kerns Palmetto Professor Endowment played an important role to this work.

**Notation**

*The following symbols are used in this paper:*

d     = particle diameter;

f     = grid frequency;

M     = size of the grid mesh;

R     = curvature radius of the particle trajectories;

$Re_g$     = grid turbulent Reynolds number;

$Re_p$     = Reynolds number based on the particle diameter and settling speed;

S     = stroke;

St     = Stokes number;

$St_K$     = Kolomogorov-based Stokes numbers;

$St_l$     = Integral scale-based Stokes numbers

$u'_{RMS}$     = root mean square of the turbulent fluctuations in the transverse direction;

v     = fluid velocity component in the normal direction;

$v'_{RMS}$     = root-mean-square of the turbulence fluctuations in the normal direction;

$v_\eta$     = Kolomogorov velocity scale;

V     = particle settling velocity in turbulent conditions;

$V_0$     = particle settling velocity in stagnant flow conditions;

x     = transverse direction (horizontal);

y     = normal direction (vertical);

$\epsilon$     = average rate of dissipation;

$\eta$     = Kolmogorov length scale;

$\mu$     = dynamic viscosity of the fluid;

ν  = kinematic viscosity of the fluid;

$\rho_p$  = particle density;

$\sigma_p$  = standard deviation of particle vertical velocity;

$\tau_f$  = characteristic fluid time scale (based on the integral length scale);

$\tau_k$  = Kolmogorov time scale;

$\tau_p$  = characteristic particle time scale;

$\mathcal{L}$  = transverse length scale (integral scale);

**Figure Captions**

**Fig. 1**: Schematic description of the experimental setup which consists of a grid turbulence tank, particle image velocimetry setup, and 2D particle tracking setup. Dimensions on the schematic are provided in centimeters.

**Fig. 2**: (a) Magnified images of the 71 μm synthetic particles, (b) magnified image of the 1400 μm natural sand particle, (c) sample size distribution from the laser diffraction size analyzer for the 261 μm natural sand particles.

**Fig. 3**: An example of time-lapse trajectories of 35 particles with a 291 μm mean diameter at 5 Hz grid oscillations.

**Fig. 4**: Instantaneous velocities in the tank at 4 Hz grid oscillations. The vectors depict the velocity as obtained by PIV and vorticity is characterized by color ($s^{-1}$).

**Fig. 5**: Vertical profiles of the RMS of turbulent fluctuations ($u'_{RMS}$) for various grid frequencies (see legend). The field of view was located ~14 cm above the bottom of the tank.

**Fig. 6**: Average $u'_{RMS}$ versus frequency.

**Fig. 7**: Auto-correlation of the vertical velocity in the horizontal direction. These curves allowed for the calculation of integral length scale ($\mathcal{L}$) as the area under this curve.

**Fig. 8**: Histograms of the particles' u and v velocities in stagnant water and turbulent conditions (see subtitles) for (left) the 261 μm natural sand and (right) the 71 μm synthetic sand.

**Fig. 9**: Settling velocities in stagnant and turbulent flow conditions versus particle size. The empirical curves of Dietrich (1982) for particle settling velocity in stagnant water are also shown for reference.

**Fig. 10**: Normalized settling velocity of particles versus Stokes number based on the integral time scales.

**Fig. 11**: Normalized standard deviation of particle settling velocity versus Stokes number based on the integral time scale.

**Fig. 12**: Normalized settling velocity of particles versus Stokes number based on Kolmogorov time scales.

**Fig. 13**: Linear combination of POD modes 3 and 4 and particle trajectories for 261 μm natural sand at 3 Hz. This figure highlights the particles' trajectory in relation to the energetic flow patterns as extracted via POD.

**Fig. 14**: Histogram of the radius of curvature for all of the particles recorded at 5 Hz grid frequency for the 291 μm industrial sand particles.

**Fig. 15**: Average radius of curvature of the particles' trajectories and corresponding POD flow pattern cell size. The smallest cells (highest modes) represented in black and largest

37    cells (lowest modes) represented in red. Dotted vertical lines are provided as a reference

38    for the range of integral length scales.

**Table 1**: Particle characteristics

| Material properties | Natural Sand | | | Industrial Sand 1 | | | Industrial Sand 2 | | Industrial Sand 3 | Synthetic |
| --- | --- | --- | --- | --- | --- | --- | --- | --- | --- | --- |
| | | | | Large | Medium | Small | Large | Small | Large | Small |
| Diameter (μm) $d_{50}$ | 1400 | 621 | 261 | 146 | 990 | 831 | 291 | 147 | 109 | 95 | 97 | 71 |
| Density (kg/m$^3$), ρ | 1650 | | | | 2640 | | | 3970 | | 2200 | 1440 |
| Relaxation time (ms), $\tau_p$ | 180.4 | 35.3 | 6.2 | 2 | 143.7 | 101.3 | 12.4 | 4.7 | 2.6 | 1.6 | 0.7 | 0.4 |
| Terminal Velocity (cm/s), $V_0$ | 8.8 | 6.6 | 3.2 | 1.3 | 8.2 | 7.2 | 2.2 | 2.0 | 0.8 | 0.5 | 0.8 | 0.5 |
| Particle Reynolds number, $Re_p$ | 123 | 41 | 8 | 2 | 82 | 60 | 6 | 3 | 1 | 0.5 | 0.8 | 0.4 |

**Table 2**: Flow characteristics

| Grid frequency $f$ (Hz) | $u'_{RMS}$ (cm/s) | Kolmogorov Scales | | | Dissipation ($\varepsilon$) (cm$^2$/s$^3$) | Integral length ($L$) (cm) | Re$_g$ |
|---|---|---|---|---|---|---|---|
| | | $\eta$ (cm) | $u_\eta$ (cm/s) | $\tau_k$ (s) | | | |
| 2 | 0.57 | 0.055 | 0.18 | 0.304 | 0.1187 | 3.6 | 5166 |
| 3 | 0.71 | 0.046 | 0.22 | 0.217 | 0.2492 | 3.9 | 7749 |
| 4 | 0.91 | 0.044 | 0.24 | 0.204 | 0.3966 | 3.8 | 10332 |
| 5 | 1.44 | 0.029 | 0.35 | 0.083 | 1.5455 | 4.1 | 12915 |
| 6 | 1.46 | 0.032 | 0.32 | 0.105 | 1.1355 | 4.2 | 15498 |
| 7 | 2.25 | 0.022 | 0.45 | 0.050 | 4.5712 | 3.7 | 18081 |

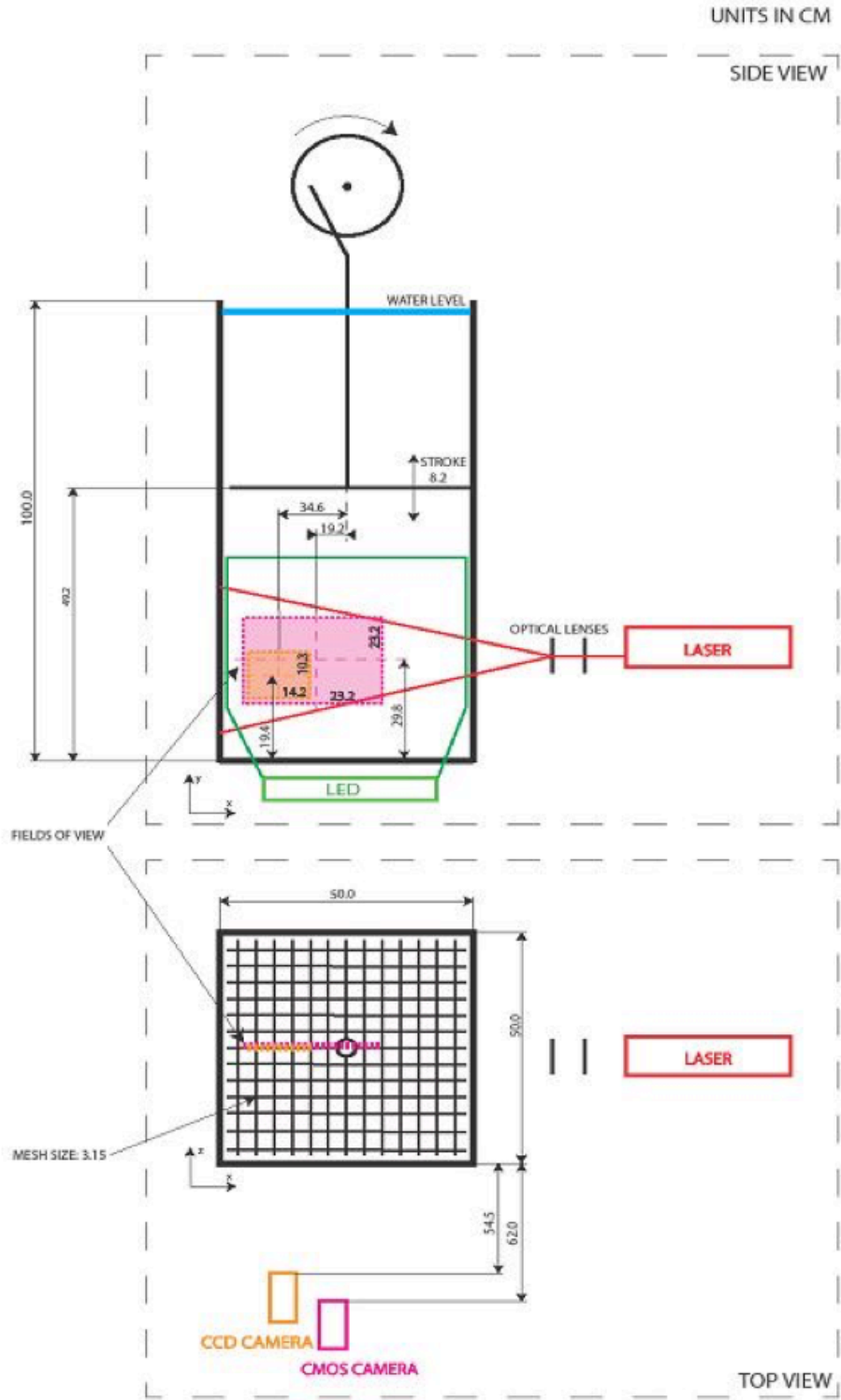

**Fig. 1**: Schematic description of the experimental setup which consists of a grid turbulence tank, particle image velocimetry setup, and 2D particle tracking setup. Dimensions on the schematic are provided in centimeters.

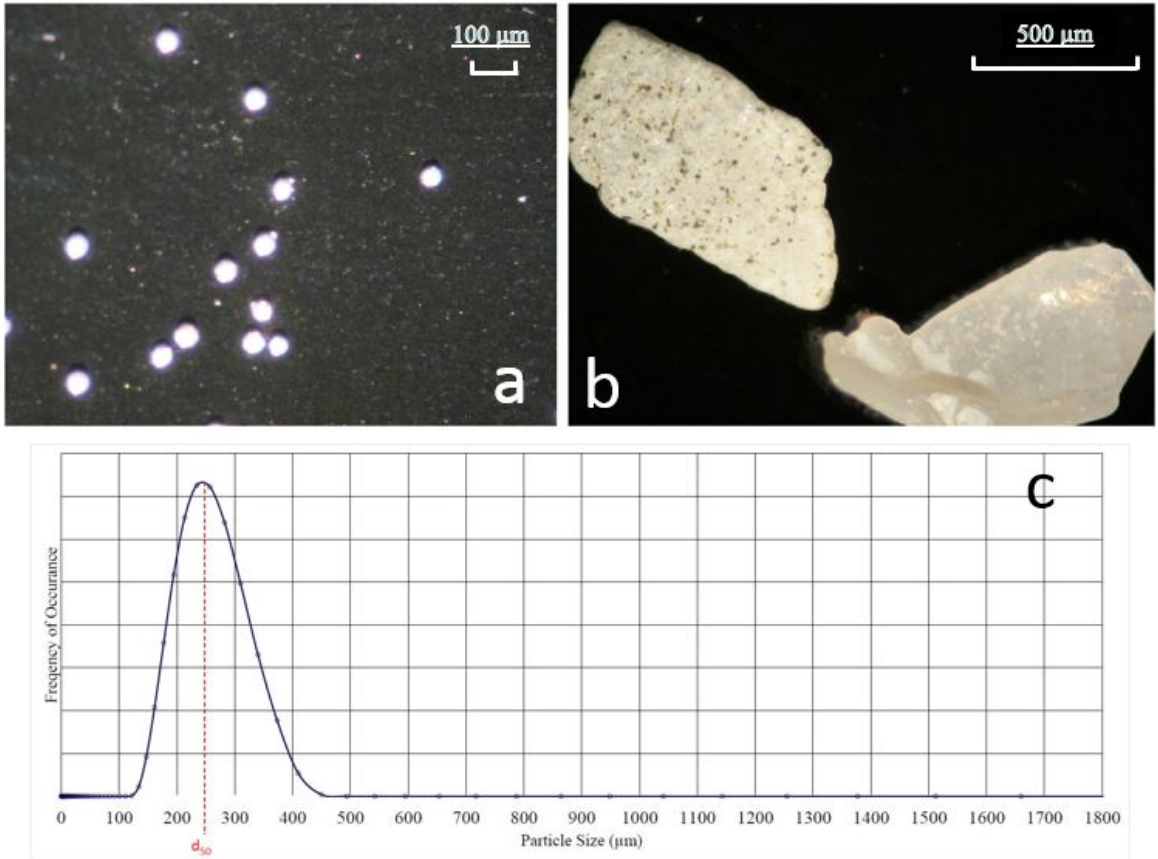

**Fig. 2**: (a) Magnified images of the 71 μm synthetic particles, (b) magnified image of the 1400-μm natural sand particle, (c) sample size distribution from the laser diffraction size analyzer for the 261 μm natural sand particles.

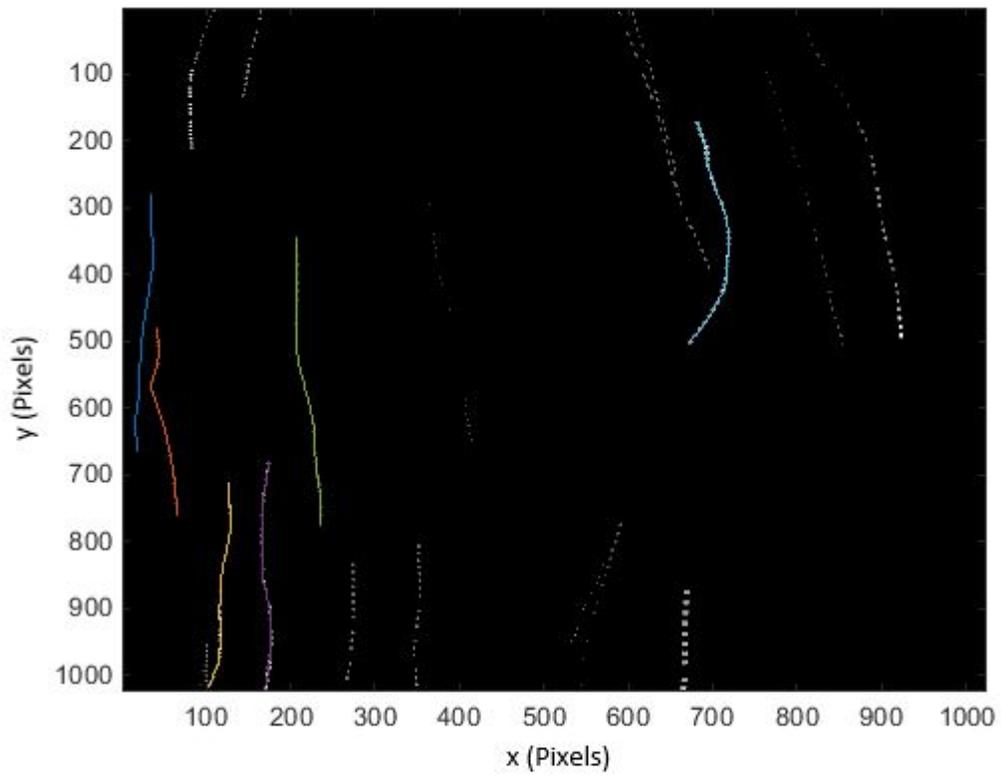

**Fig. 3**: An example of time-lapse trajectories of 35 particles with a 291 μm mean diameter at 5 Hz grid oscillations.

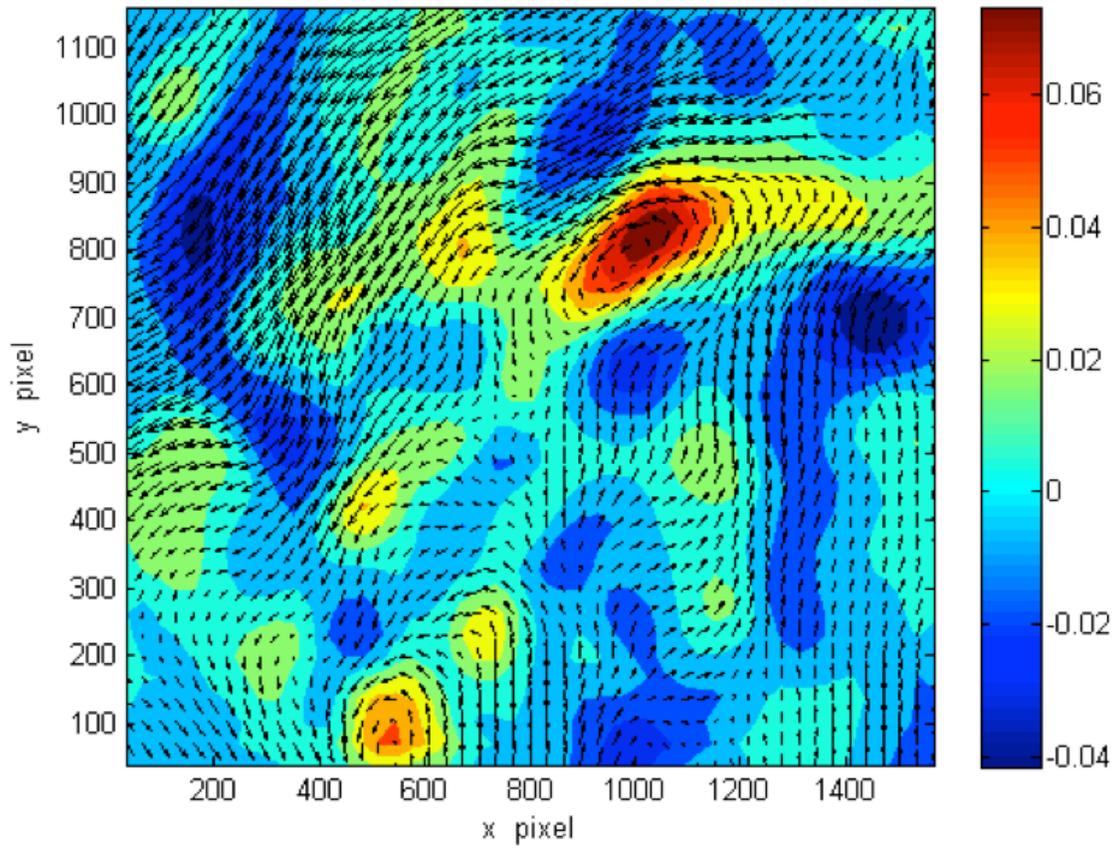

**Fig. 4**: Instantaneous velocities in the tank at 4 Hz grid oscillations. The vectors depict the velocity as obtained by PIV and vorticity is characterized by color (s$^{-1}$).

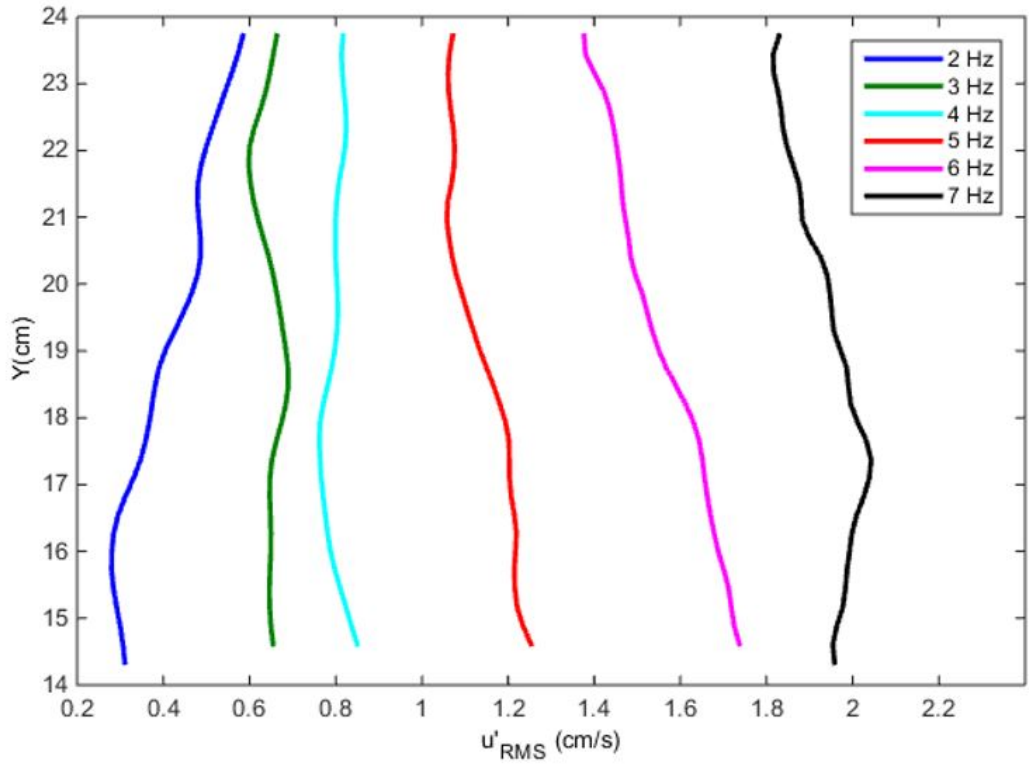

**Fig. 5**: Vertical profiles of the RMS of turbulent fluctuations (u'$_{RMS}$) for various grid frequencies (see legend). The field of view was located ~14 cm above the bottom of the tank.

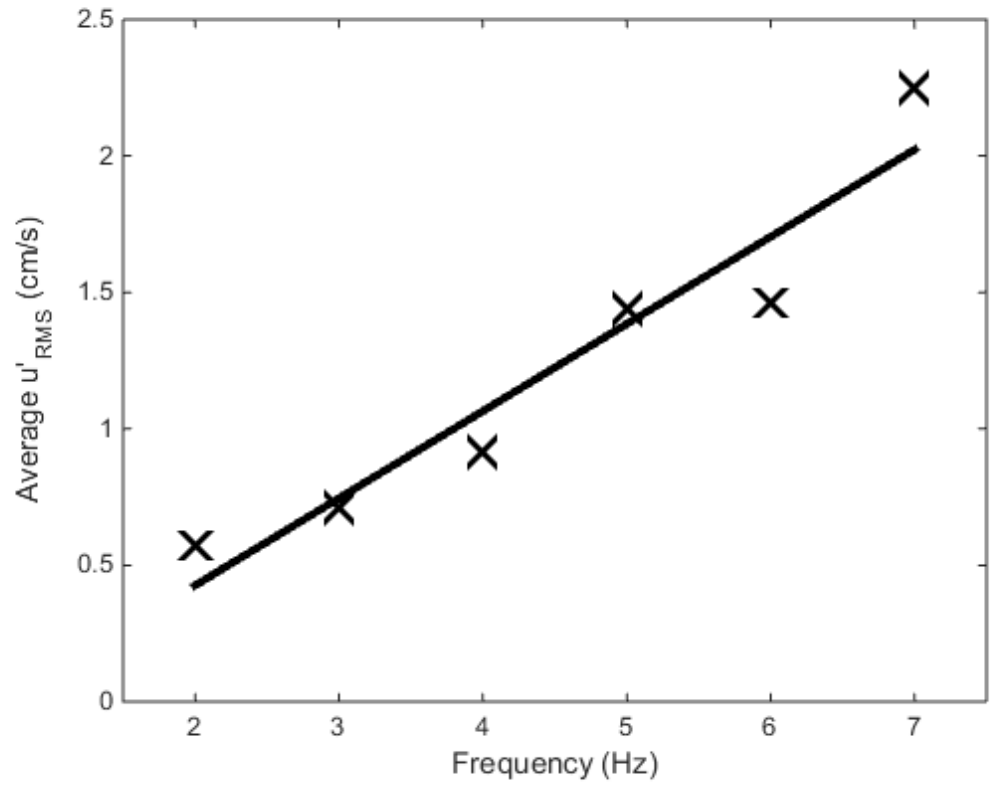

**Fig. 6**: Average u'$_{RMS}$ versus frequency.

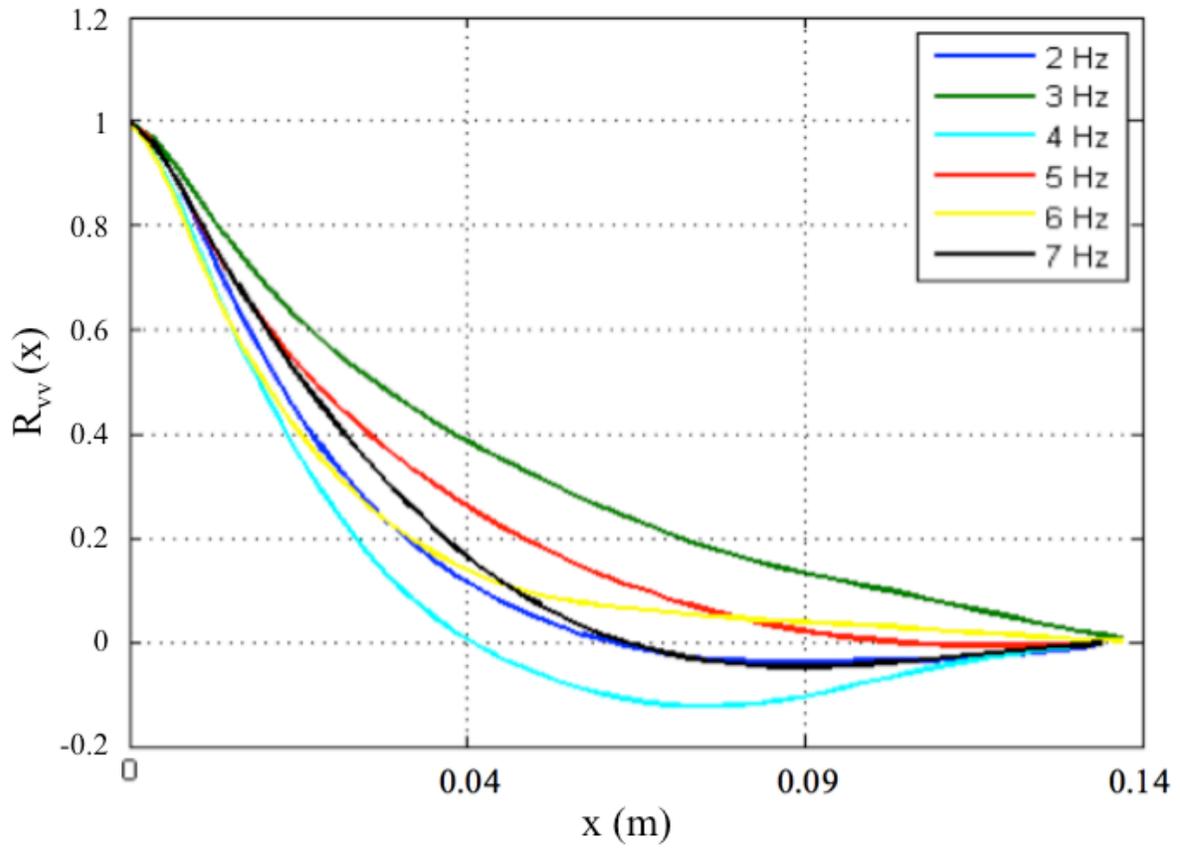

**Fig. 7**: Auto-correlation of the vertical velocity in the horizontal direction. These curves allowed for the calculation of integral length scale ($\mathcal{L}$) as the area under this curve.

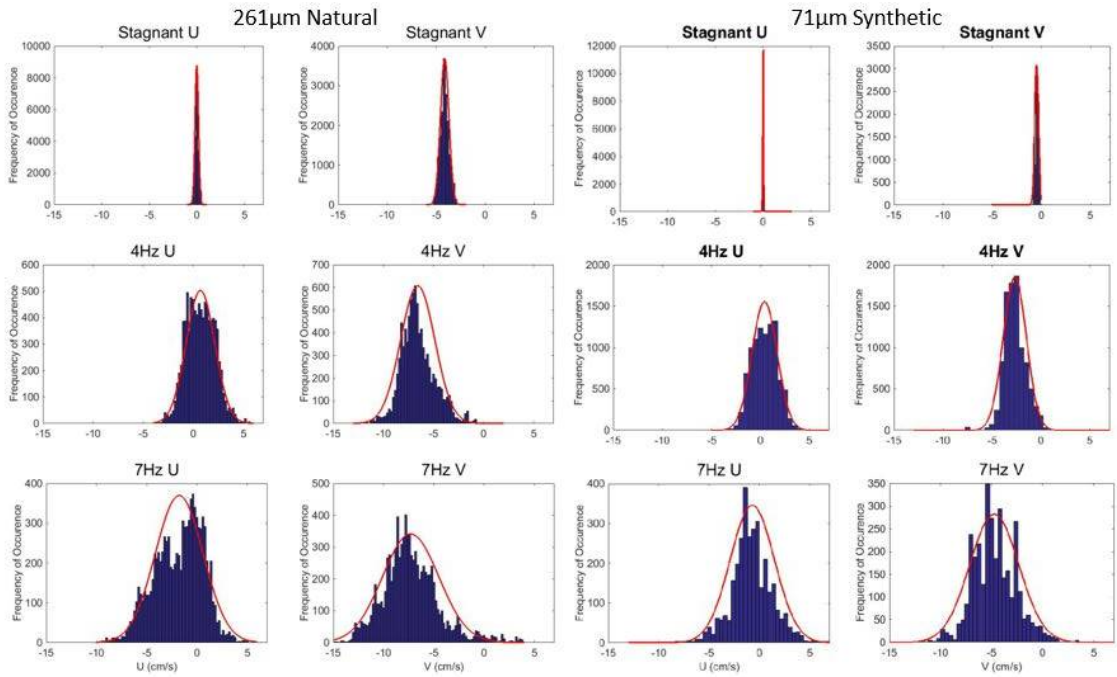

**Fig. 8**: Histograms of the particles' u and v velocities in stagnant water and turbulent conditions (see subtitles) for (left) the 261 μm natural sand and (right) the 71 μm synthetic sand.

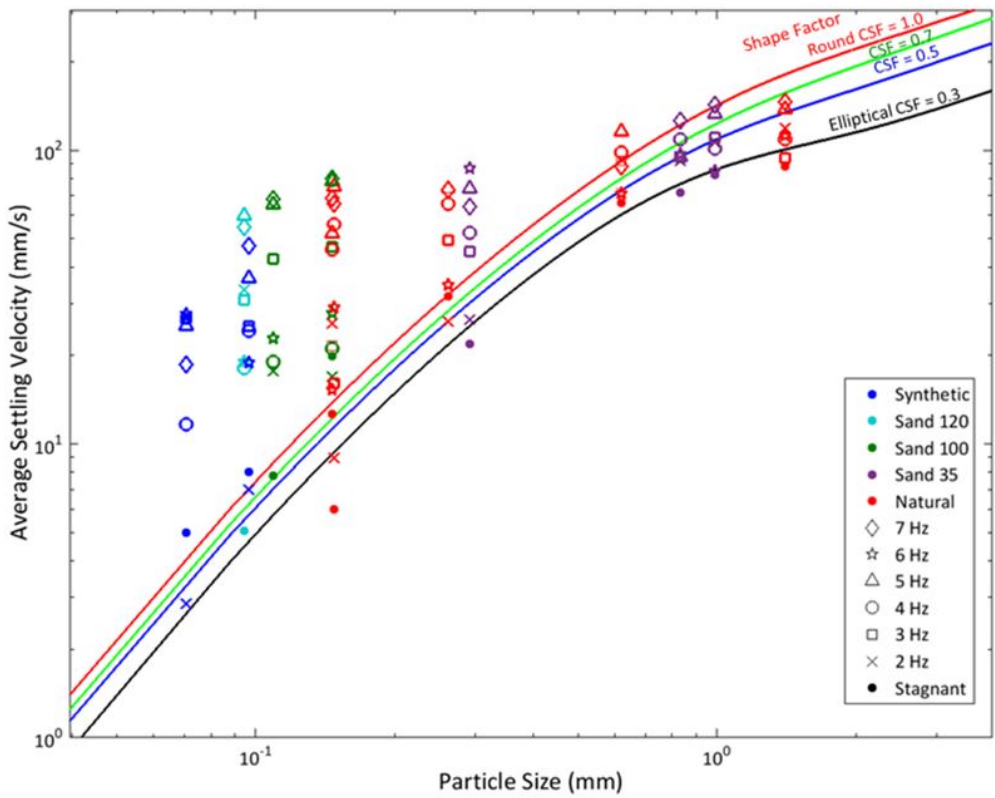

**Fig. 9**: Settling velocities in stagnant and turbulent flow conditions versus particle size. The empirical curves of Dietrich (1982) for particle settling velocity in stagnant water are also shown for reference.

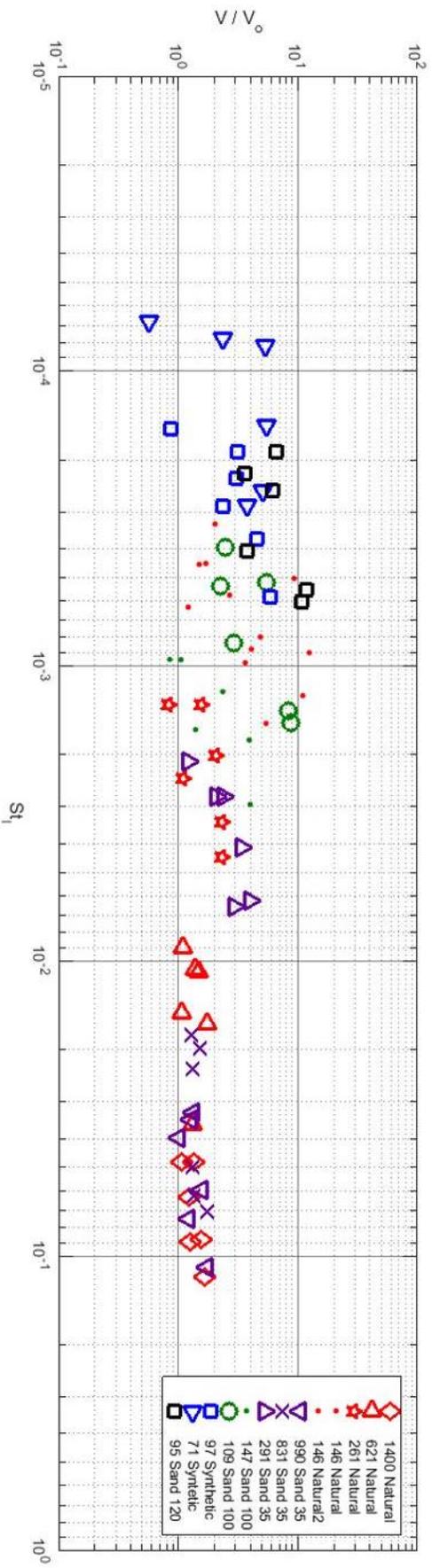

**Fig. 10**: Normalized settling velocity of particles versus Stokes number based on the integral time scales.

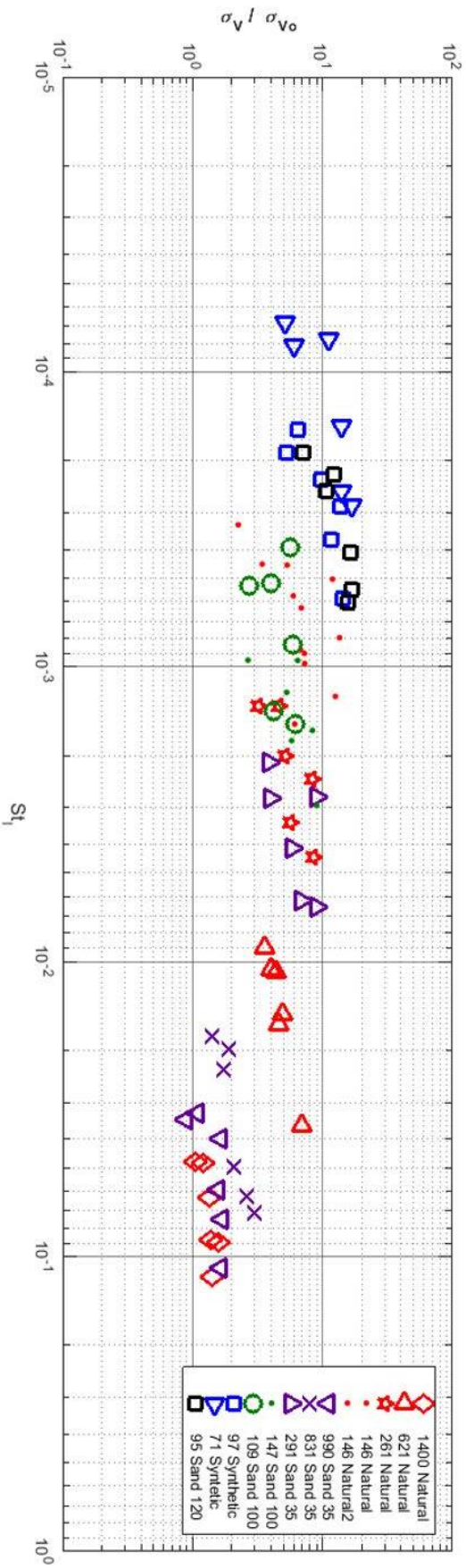

**Fig. 11**: Normalized standard deviation of particle settling velocity versus Stokes number based on the integral time scale.

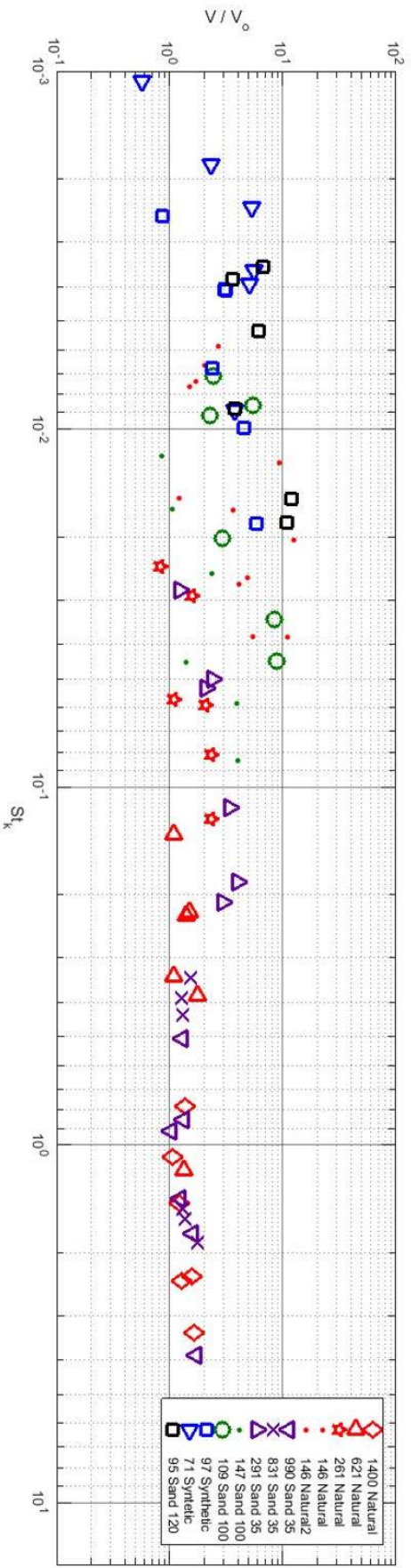

**Fig. 12**: Normalized settling velocity of particles versus Stokes number based on Kolmogorov time scales.

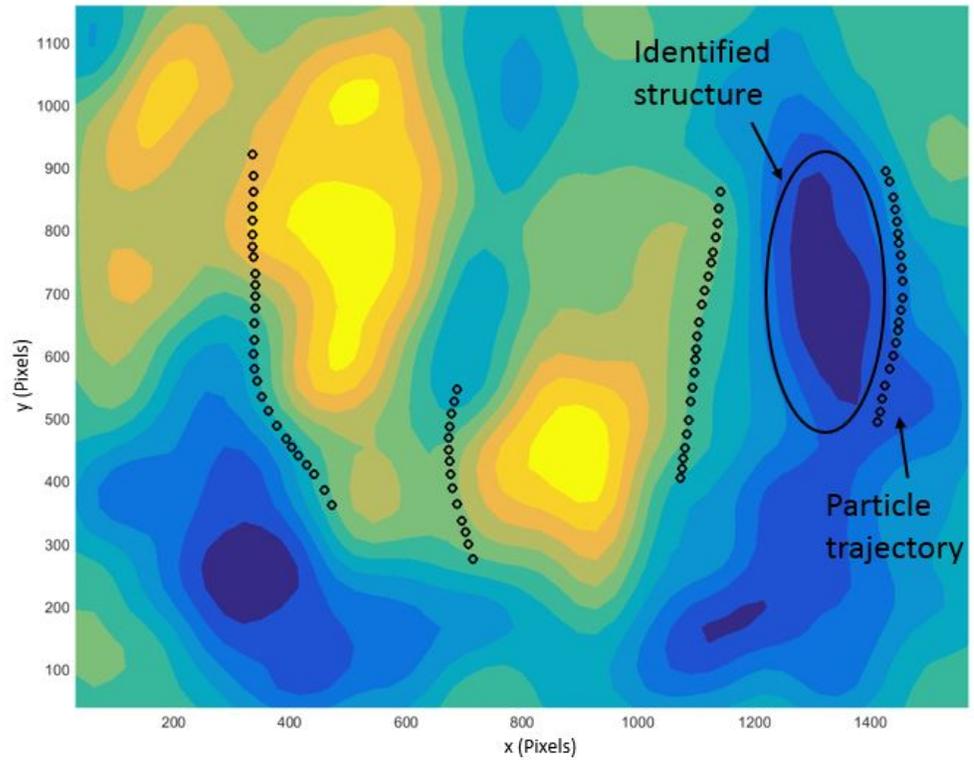

**Fig. 13**: Linear combination of POD modes 3 and 4 and particle trajectories for 261 μm natural s 3 Hz. This figure highlights the particles' trajectory in relation to the energetic flow patte extracted via POD.

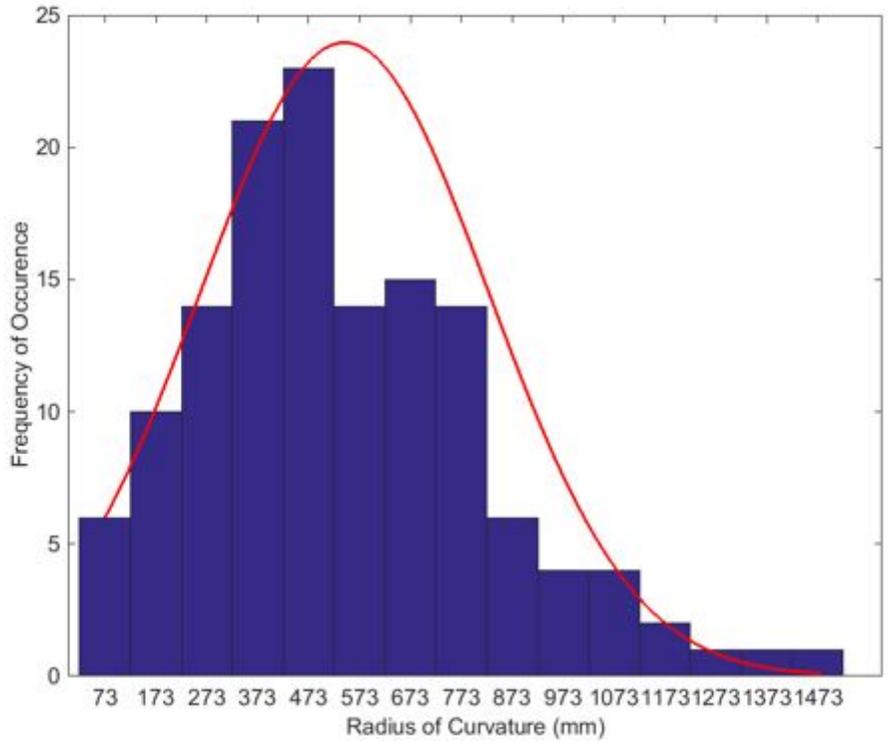

**Fig. 14**: Histogram of the radius of curvature for all of the particles recorded at 5 Hz grid frequency for the 291 μm industrial sand particles.

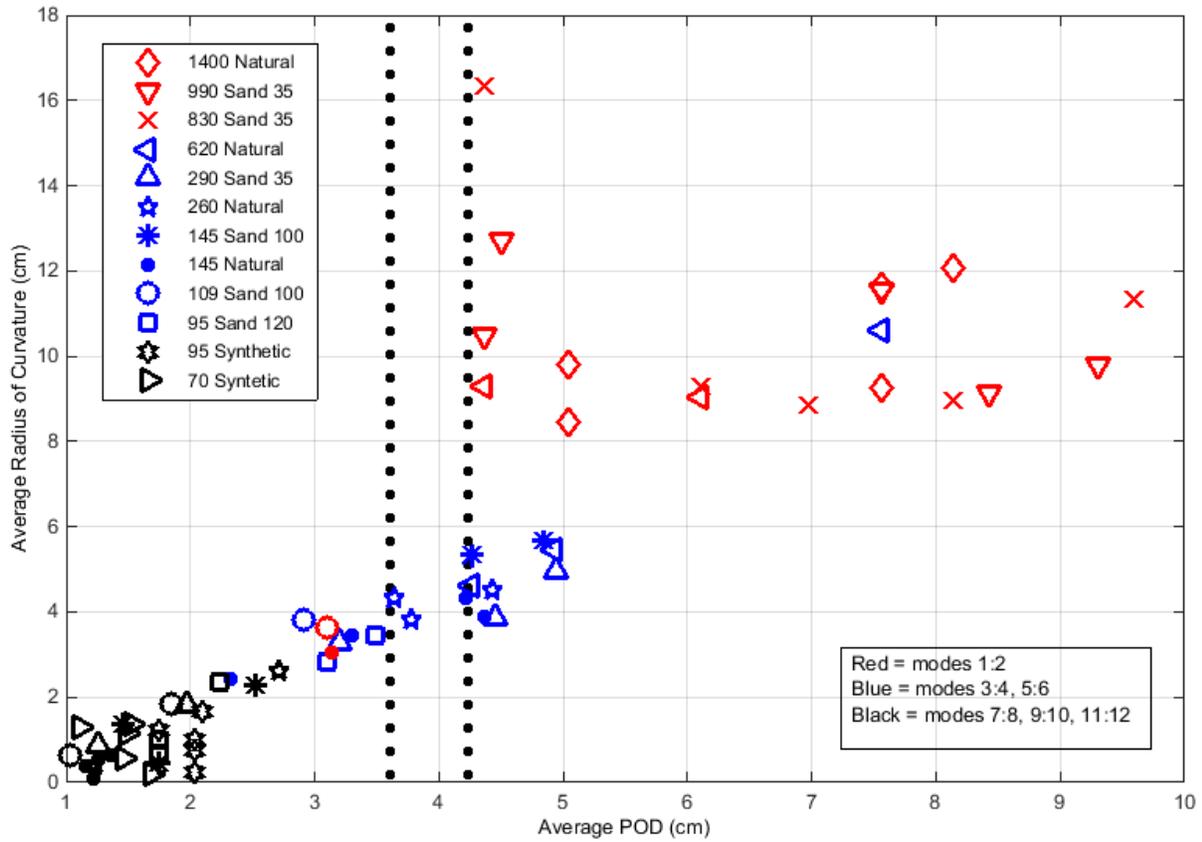

**Fig. 15**: Average radius of curvature of the particles' trajectories and corresponding POD flow pattern cell size. The smallest cells (highest modes) represented in black and largest cells (lowest modes) represented in red. Dotted vertical lines are provided as a reference for the range of integral length scales.